\documentclass[aps,prl,reprint,superscriptaddress]{revtex4-2}

\usepackage{amsmath}
\usepackage{amssymb}
\usepackage{xcolor}
\DeclareMathOperator{\sgn}{sgn}

\usepackage{graphicx}
\usepackage{dcolumn}
\usepackage{bm}

\begin{document}

\preprint{APS/123-QED}

\title{New Massless Spectra from Cosmic String Cusps}

\author{Amelia Drew}
\affiliation{The Abdus Salam ICTP, Strada Costiera 11, 34151, Trieste, Italy}
\affiliation{Institute for Fundamental Physics of the Universe, Via Beirut 2, 34151 Trieste, Italy}
\affiliation{Department of Applied Mathematics and Theoretical Physics,
University of Cambridge, Wilberforce Road, Cambridge CB3 0WA, United Kingdom}
\affiliation{Homerton College, Hills Road, Cambridge CB2 8PH, United Kingdom}

\author{Ivan Rybak}
\affiliation{Cosmology, Gravity, and Astroparticle Physics Group, Center for Theoretical Physics of the Universe, Institute for Basic Science (IBS), Daejeon, 34126, Korea}
\affiliation{CAPA \& Departamento de F\'{\i}sica Te\'{o}rica, Universidad de Zaragoza, Pedro Cerbuna, 12, Zaragoza, 50009}

\date{\today}

\begin{abstract}
In the standard picture of cosmic strings, cusps are generic features of Nambu–Goto loops where the string momentarily reaches the speed of light. They have a characteristic sharp profile, following $y \sim x^{2/3}$ in the $(x,y)$ plane, and produce strong gravitational-wave (GW) bursts with waveform $\tilde\kappa(\omega) \propto \omega^{-4/3}$, making them key targets for GW searches. However, under certain conditions, this generic picture can differ. We identify novel cusp solutions with different, including smooth, shapes, and compute their massless GW and axion spectra. We derive a general expression for all possible Nambu-Goto cusp spectra with the asymptotic form $\tilde\kappa(\omega) \propto \omega^{-n/(2n-1)} \omega^{-m/(2m-1)}$ where $n, m\geq 2$. We calculate the effect on the waveform of realistic corrections to the Nambu-Goto description, such as those from backreaction and finite string width, which introduce a high frequency cutoff. 
\end{abstract}

\maketitle


\textit{Introduction}---Cosmic strings are a class of topological defect that arise in high-energy physics models with spontaneously broken symmetry, where the vacuum manifold $\mathcal{M}$ has a non-trivial fundamental group $\pi_1(\mathcal{M}) \neq I$ \cite{Kibble:1976sj,Kibble:1999yk}. If detected, they would provide direct evidence of new physics beyond the Standard Model \cite{Hindmarsh:1994re,Vilenkin:2000jqa}. Networks of such strings are expected to arise cosmologically, comprising long strings that stretch across Hubble patches and loops that oscillate, radiate and collapse over time. Their predicted gravitational-wave (GW) signatures \cite{Damour:2001bk} have made them a target for searches by LIGO-Virgo-KAGRA
\cite{LIGOScientific:2017ikf,LIGOScientific:2021nrg}, as well as pulsar timing experiments, e.g. \cite{NANOGrav:2023hvm, EuropeanPulsarTimingArray:2023lqe}. The spectrum of axion (Nambu-Goldstone boson) radiation from axion strings, crucial for calculating the axion mass for a post-inflation symmetry breaking, has also been a longstanding topic of debate \cite{Gorghetto:2018myk,Gorghetto:2020qws,Buschmann:2021sdq,Kim:2024wku,Saikawa:2024bta,Correia:2024cpk,Correia:2025nns}.

Much of the physical behaviour of cosmic strings can be effectively modeled using the Nambu-Goto action, either in isolation or coupled to additional fields (e.g. a Kalb-Ramond field \cite{Davis:1988rw}). Within this framework, it is well established that certain generic features, such as cusps and kinks, naturally emerge on oscillating string loops due to their periodic dynamics. These features source strong, localised bursts of GWs \cite{Burden:1985md,Garfinkle:1987yw,Battye:1993jv,Vilenkin:1986ku,Damour:2000wa, Damour:2001bk} and produce a high frequency tail in the stochastic GW background, making them a focus of GW string searches \cite{LIGOScientific:2019ppi, LIGOScientific:2021nrg}. 


In this work, we identify novel Nambu-Goto cusp configurations not previously identified in the literature. We compute their shapes and massless (GW and axion) waveforms, and provide explicit examples generated from the collision of two travelling waves on a long string worldsheet \cite{Vachaspati:1990sk, Olum:1998ag, Drew:2023ptp}. We also calculate the effect on cusp waveforms of small deviations from the Nambu-Goto approximation, relevant, for example, for finite width effects or backreaction in field theory strings. These new waveforms could be used to improve and broaden experimental GW cusp searches. 





\textit{Review of Nambu-Goto Dynamics}---The classical evolution of a one-dimensional cosmic string can be modelled using the Nambu-Goto effective action \cite{Vilenkin:2000jqa, Zwiebach:2004tj}
\begin{equation}\label{NG}
    S = -\mu\int{\sqrt{- \gamma}}\,\mathrm{d}^2\zeta\,,
\end{equation}
where $\mu$ is the string tension, $\gamma \equiv \mathrm{det}(\gamma^{ab})$ is the determinant of the worldsheet metric $\gamma_{ab}=X^{\mu}_{,a} X^{\nu}_{,b} g_{\mu \nu}$, $a,b = 0,1$ label worldsheet parameters $\zeta^a \equiv (\tau,\sigma)$, $g_{\mu \nu}$ is the spacetime metric, $X^\mu(\zeta^a)$ is the position of the string core and $X^{\mu}_{,a} \equiv \frac{\partial X^{\mu}}{\partial \zeta^a}$. In a Minkowski background $\eta_{\mu\nu}$ in the conformal gauge (for which $\gamma_{ab} = \sqrt{-\gamma}\,\eta_{a b}$) the equations of motion reduce to wave equations
\begin{equation}\label{wave}
    \ddot{X}^\mu - X''^\mu = 0.
\end{equation}
We can write the general solution 
\begin{equation}
    \label{EqOfMot}
    X^0 = \tau, \qquad
    \textbf{X}(\tau,\sigma) = \frac{1}{2} \left( \textbf{X}_+(\sigma_+) + \textbf{X}_-(\sigma_-) \right),
\end{equation}
where $\sigma_{\pm} = \tau \pm \sigma $ and $\textbf{X}_{\pm}(\sigma_{\pm})$ denote components moving in opposite directions along the string, usually referred to as left- (right-) moving modes. The gauge conditions lead to the following condition for each mode:
\begin{equation}
\label{UnitVectors}
|\,\textbf{X}^{(1)}_{\pm}(\sigma_{\pm})\,| = 1,   
\end{equation}
where $\textbf{X}^{(n)}_{\pm} \equiv \frac{d^n \textbf{X}_{\pm}(\sigma_{\pm})}{d \sigma_{\pm}^n}$. The evolution of the string can therefore be described by trajectories of $\textbf{X}^{(1)}_{\pm}$ on a 2-dimensional `Kibble-Turok' unit sphere \cite{Kibble:1982cb, Turok:1984cn}. 

In this framework, certain points on a string can reach the speed of light $\dot{\textbf{X}}_{\rm c}^2 = 1$, forming a `cusp'. This occurs whenever
\begin{equation}
\label{CuspCond}
\textbf{X}^{(1)}_{\rm c +} = \textbf{X}^{(1)}_{\rm c -},
\end{equation}
corresponding to a point of intersection of the trajectories traced by $\textbf{X}^{(1)}_{\pm}$, which arise generically in string loops without discontinuities in $\textbf{X}^{(1)}_{\pm}$ (i.e., in the absence of kinks) \cite{Kibble:1982cb, Turok:1984cn}. We can also construct cusp solutions on long strings (i.e. non-periodic) by choosing left- and right-moving modes appropriately. Long string configurations have been modelled analytically \cite{Garfinkle:1987yw, Garfinkle:1990jq, Vachaspati:1990sk} and also studied numerically \cite{Blanco-Pillado:1998tyu, Drew:2019mzc, Drew:2022iqz, Drew:2023ptp}. 

\textit{Shapes of Cusps}---To calculate the shape of a cusp in real space, we Taylor expand $\mathbf{X}$ around the cusp point $\textbf{X}_{\rm c}$ (see e.g. \cite{Vilenkin:2000jqa, Zwiebach:2004tj}). Separating into $\mathbf{X_\pm}$ and expanding with respect to $\sigma_{\pm}$, 
\begin{equation}
\begin{gathered}
    \label{CuspTaylor}
    \textbf{X}_{\pm} (\sigma_{\pm}) = \textbf{X}_{\rm c \pm}  + \sum^{\infty}_{n=1}\frac{\sigma^n_{\pm}}{n!} \textbf{X}^{(n)}_{\rm c \pm}\,,
\end{gathered}
\end{equation}
where the cusp occurs at $\sigma_{\rm c \pm} = 0$. 



In the most general case, terms up to $\textbf{X}^{(3)}_{\pm}$ in \eqref{CuspTaylor} are important, and we do not have any conditions on the higher derivatives of the string core position; $\mathbf{X}_{\mathrm{c}\pm}^{(n)} \neq 0$ for $n \geq 2 $, and in general these higher derivatives will be smooth functions \cite{Blanco-Pillado:1998tyu}. Taylor expanding $\textbf{X}(0,\sigma)$ \eqref{EqOfMot} around the cusp point, we obtain
\begin{align}
    \label{ShapeGeneric}
    \textbf{X}(0,\sigma) &=  \frac{\sigma^2}{4}\left( \textbf{X}^{(2)}_{\rm c+} + \textbf{X}^{(2)}_{\rm c-} \right) + \frac{\sigma^3}{12} \left( \textbf{X}^{(3)}_{\rm c+} - \textbf{X}^{(3)}_{\rm c-}  \right) + \mathcal{O}(\sigma^4)
\end{align}
where the cusp occurs at $\mathbf{X}_{\mathrm{c}} = \mathbf{0}$ and the term linear in $\sigma$ vanishes due to the cusp condition \eqref{CuspCond}.

\begin{figure}[h!]
	\centering \includegraphics[width=0.48\textwidth, angle=0, trim = 0.7cm 0 0 0, clip]{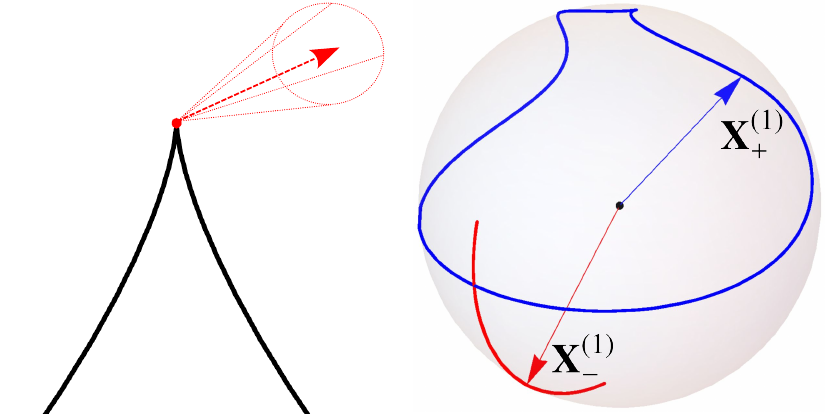}	\\
    \hspace{.3cm}
    \includegraphics[width=0.48\textwidth, angle=0]{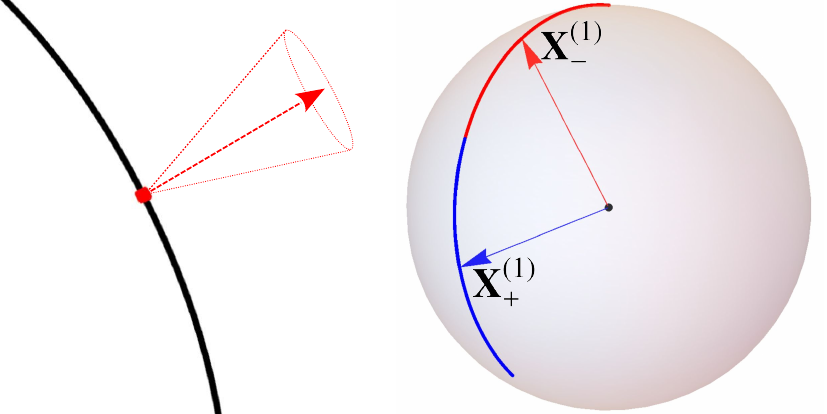}\\
\caption{Top: The $y\sim x^{2/3}$ shape of the generic cusp in physical space, where the dot represents a point moving at the speed of light, and corresponding Kibble-Turok sphere. The direction of the radiation is schematically indicated. Bottom: The shape of a smooth cusp in physical space, where the discontinuous function $\textbf{X}_{\rm c}^{(2)}$ smooths out the shape, and corresponding Kibble-Turok sphere.} 
	\label{Fig:Cusp1}%
\end{figure}

We realise this configuration in real space by aligning components of \eqref{ShapeGeneric} with appropriate Cartesian axes. Differentiating the condition \eqref{UnitVectors} with respect to the worldsheet parameters, each mode obeys 
\begin{equation}
    \label{GenericCuspDiff}
    \textbf{X}^{(1)}_{\pm} \cdot \textbf{X}^{(2)}_{\pm}=0, \qquad \textbf{X}^{(3)}_{\pm} \cdot \textbf{X}^{(1)}_{\pm}= -\,\left(\textbf{X}^{(2)}_{\pm}\right)^2\,.
\end{equation}
It can be shown (e.g. \cite{Zwiebach:2004tj}) that the cusp shape is given by $\textbf{X} \sim \left\{\sigma^3,  \sigma^{2},0 \right\}$ for small $\sigma$, i.e. the string traces \begin{equation}y \sim x^{2/3}\end{equation} in the $(x,y)$ plane near the cusp point, as shown in Figure \ref{Fig:Cusp1}. Since $\textbf{X}^{\prime}=0$ and $\textbf{X}^{\prime \prime} \neq 0$ and recalling that the curvature of a parametrised curve is given by $\kappa(\sigma) = || \mathbf{X}^{\prime} \times \mathbf{X}^{\prime \prime} ||\,/\, || \mathbf{X}^{\prime} ||^3$, the cusp point is characterised by an infinite curvature on the worldsheet.

If we consider cases where certain terms in \eqref{CuspTaylor} vanish, or consider different smoothness conditions, we obtain cusps with different shapes. We first consider the case where $\textbf{X}_{\pm}^{(n)} (\sigma_{\pm})$ changes sign at the cusp point, i.e.
\begin{equation}
\label{Disc2Der}
\lim_{\sigma \rightarrow  \sigma^+_{\pm \rm c}} \textbf{X}_{\pm}^{(n)} (\sigma_{\pm}) = - \lim_{\sigma \rightarrow  \sigma^-_{\pm \rm c}} \textbf{X}_{\pm}^{(n)} (\sigma_{\pm}) \neq 0 \,.
\end{equation}
Taking $n=2$ as an example, the expansion around the cusp point takes the form
\begin{equation}
\label{ShapeSmooth}
\begin{aligned}
    \textbf{X}(0,\sigma) = \sgn(\sigma) \frac{\sigma^2}{4}&\left( \textbf{X}^{(2)}_{\rm c+} + \textbf{X}^{(2)}_{\rm c-} \right) + \\ &\frac{\sigma^3}{12}\left( \textbf{X}^{(3)}_{\rm c+} - \textbf{X}^{(3)}_{\rm c-} \right) + \mathcal{O}(\sigma^4) \,.
\end{aligned}
\end{equation}
In the limit where $\sigma \rightarrow \sigma^+_c$, the shape is given by $\textbf{X} \sim \left\{\sigma^3,  \sigma^{2},0 \right\}$ and for $\sigma \rightarrow \sigma^-_c$, by $\textbf{X} \sim \left\{\sigma^3, - \sigma^{2},0 \right\}$. Remarkably, \eqref{Disc2Der} produces a smooth cusp as shown in Figure \ref{Fig:Cusp1}, with 
\begin{equation}
|\,y\,| \sim x^{2/3}\,,
\end{equation} in contrast to \eqref{GenericCuspDiff}, where the curvature at the cusp point diverges. We emphasise that all formal cusp criteria are satisfied. Physically, \eqref{ShapeSmooth} arises when the string reaches zero acceleration and then begins accelerating in the opposite direction. This follows from the observation that $\ddot{\textbf{X}}(\tau,0) \sim \sgn(\tau)\left(\textbf{X}^{(2)}_{{\rm c}+} - \textbf{X}^{(2)}_{{\rm c}-}\right)$, as encountered in \cite{Drew:2023ptp}. 



Another example of a non-generic cusp shape arises when $\textbf{X}^{(p)}_{{\rm c} \pm}$ vanish for all $1 < p < n$, where $n$ is the smallest integer for which $\textbf{X}^{(n)}_{{\rm c} \pm} \neq 0$. Here, one must examine higher-order derivative constraints obtained by differentiating \eqref{UnitVectors}. The conditions on the fourth- and fifth-order derivatives are 
\begin{align}
    \textbf{X}^{(4)}_{\pm} \cdot \textbf{X}^{(1)}_{\pm} &= -3\,\textbf{X}^{(2)}_{\pm} \cdot \textbf{X}^{(3)}_{\pm}, \\ 
    \textbf{X}^{(5)}_{\pm} \cdot \textbf{X}^{(1)}_{\pm} &= -3\,\left(\textbf{X}^{(3)}_{\pm}\right)^2 - 4\,\textbf{X}^{(2)}_{\pm} \cdot \textbf{X}^{(4)}_{\pm}\,.
\end{align}
Here, we can generically (although not always) choose a coordinate system in which the $n$-th and $(n$$+$$1)$-th derivatives lie in the $(x,y)$ plane, producing a cusp shape
\begin{equation}\label{genericnewshape}
y \sim x^{n/(n+1)}\,.
\end{equation} 





\begin{figure}
\begin{center}
\includegraphics[scale=0.47]{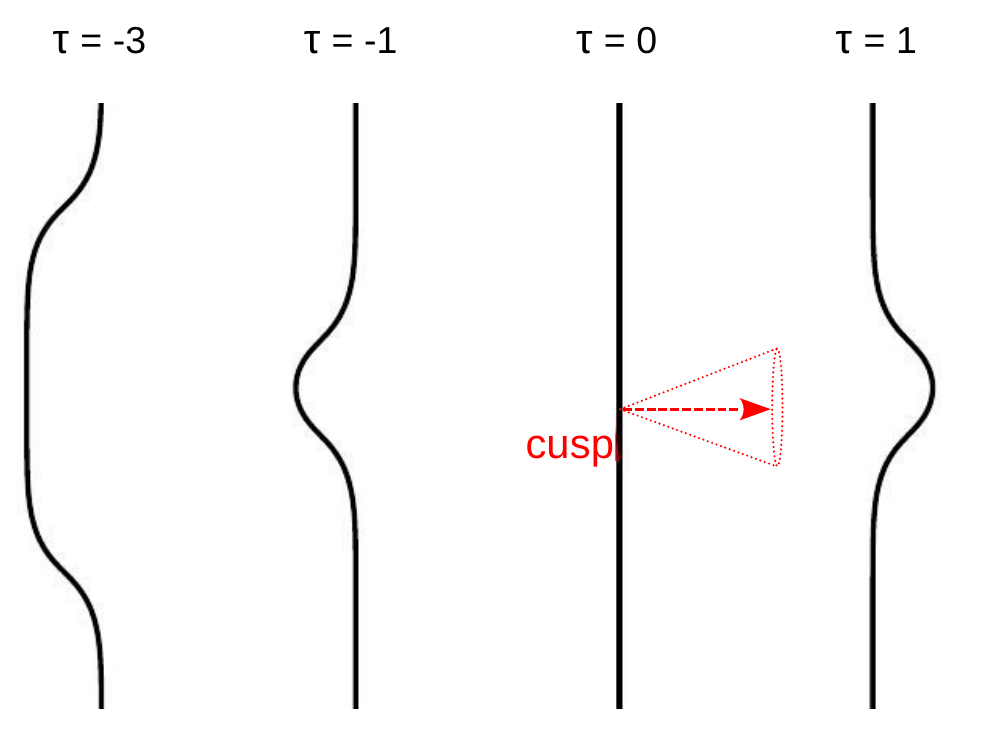}
\caption{\label{fig:CrzyCusp} Cusp formation with $\textbf{X}^{(2)}_{{\rm c} \pm} = 0$ from the collision of travelling waves at $\tau=0$ with $\sigma=0$, described by \eqref{CrazyCusp}. The direction of radiation is schematically indicated.} 
\end{center}
\end{figure}

We can investigate the shapes of different cusp configurations using travelling wave solutions with specific properties 
The configuration
\begin{equation}
\begin{gathered}
    \label{CrazyCusp}
\textbf{X}_{\pm}^{(1)} = \left\{  \sin\left( \frac{\pi}{2} \text{e}^{-\alpha |\sigma_{\pm}|^{n-1}} \right), \quad \pm \cos\left( \frac{\pi}{2} \text{e}^{-\alpha |\sigma_{\pm}|^{n-1}} \right), \quad 0 \right\},
\end{gathered}
\end{equation}
describes two travelling waves whose independent propagation along the string does not produce radiation \cite{Vachaspati:1990sk}. Here, $\alpha$ is a positive constant, and we can specify the order of the lowest non-zero derivative $n$. In the far past, $\sigma_{\pm} \to -\infty$, the waves travel freely from opposite directions along the string. As they approach $\sigma_{\pm} = 0$, they begin to interact and eventually collide, forming a cusp, as can be explicitly verified from \eqref{CuspCond}. Here, the cusp forms when the string is straight, $\kappa(\sigma) = || \mathbf{X}^{\prime} \times \mathbf{X}^{\prime \prime} ||\,/\, || \mathbf{X}^{\prime} ||^3 \to 0 $; the nontrivial vectors $\textbf{X}^{(n)}_{{\rm c}\pm}$ are oriented in the same direction, so $\textbf{X}(0,\sigma)$ has only one nonvanishing component. The evolution for $n = 3$ ($\textbf{X}^{(2)}_{{\rm c} \pm} = 0$) is shown in Figure~\ref{fig:CrzyCusp}.

Since the cusp illustrated in Figure~\ref{fig:CrzyCusp} appears completely straight, one may ask whether higher-order corrections associated with the finite string thickness can contribute to the motion in the vicinity of this cusp. For a standard cusp, higher-order corrections due to the finite string thickness diverge, signaling the breakdown of the Nambu--Goto approximation. This can be seen explicitly by studying the extrinsic curvature \cite{Anderson:1997ip},
\begin{equation}
K_{A \, a b} \equiv  n^{\mu}_{A \, ,a} X^{\nu}_{,b} g_{\mu \nu},
\end{equation}
where $n_A$ are orthonormal vectors normal to the string worldsheet. By calculating the components of the extrinsic curvature explicitly, one finds that
\begin{equation}
\lim_{t,\sigma \rightarrow 0} K_{A \, a b} = 0.
\end{equation}
Consequently, the dominant, and potentially the only, correction to the Nambu-Goto dynamics for such cusps is expected to arise from the radiation of massive particles, where the vector component of the radiation can be described by the effective model proposed in \cite{Rybak:2023bky}. 


Another interesting configuration occurs when the first derivative of the left- or right-moving mode is a constant vector. For instance, if $X^{(1)}_{+} = \{1, 0, 0\}$, the cusp conditions (5) are always satisfied at some time $\tau = -\sigma$. A persistent cusp thus forms as a travelling wave. In contrast to previous examples, this cusp does not radiate~\cite{Vachaspati:1984gt}.

\textit{Massless Radiation Spectra}---Both gravitational and axion radiation can be described by wave equations with source terms. The (symmetric) trace-reversed metric perturbation $\bar{h}^{\mu \nu} \equiv h^{\mu \nu} - h^{\lambda}_{\lambda} \eta^{\mu \nu}$, where $h^{\mu \nu}$ is a linear perturbation $g_{\mu \nu} = \eta_{\mu \nu} + h_{\mu \nu}$ around a background metric $\eta_{\mu \nu}$, is sourced by the stress-energy tensor $T^{\mu\nu}(x)$ through the linearised Einstein equations
\begin{equation}
    \Box \bar{h}^{\mu \nu} = 4 \pi \kappa_g T^{\mu \nu}\,
\end{equation}
in the $(+---)$ signature, where $\kappa_{g} = 4 G$ in natural units. We can use the duality between a scalar (axion) field $\vartheta$ and an antisymmetric two-form $B^{\mu \nu}$,
\begin{equation}
    \eta \partial_\mu\vartheta = \frac{1}{2}\epsilon_{\mu\nu\lambda\rho}\partial^\nu B^{\lambda\rho}\,,
\end{equation}
where $\eta$ is the energy scale of the symmetry breaking, to write an equation for the field sourced by current $J^{\mu\nu}(x)$,
\begin{equation}
    \label{GRrad}
     \qquad \Box B^{\mu \nu} = 4 \pi \kappa_a J^{\mu \nu}\,,
\end{equation}
where $\kappa_{a}=1$ is a constant. The Fourier transformed (denoted by tilde) asymptotic waveforms in the local wave zone for gravitational waves $\kappa^{\mu\nu}$ \cite{Damour:2000wa, Damour:2001bk, Damour:2004kw} and axions $b^{\mu\nu}$ are given respectively by
\begin{align}
    \label{GrRadExpr}
    \tilde{\kappa}^{\mu \nu}(\omega, \hat{\textbf{n}}) \equiv r \tilde{\bar{h}}^{\mu \nu}(\omega, \hat{\textbf{n}}) \approx \kappa_g \text{e}^{i \omega r} \tilde{T}^{\mu \nu} (\omega, \hat{\textbf{n}}) , \\
    \tilde{b}^{\mu \nu} (\omega, \hat{\textbf{n}}) \equiv r \tilde{B}^{\mu \nu} (\omega, \hat{\textbf{n}}) \approx \kappa_a \text{e}^{i \omega r} \tilde{J}^{\mu \nu}(\omega, \hat{\textbf{n}}),
\end{align}
where the distance to the source $r$ is much bigger than the size of the source and $\hat{\textbf{n}}$ is a unit vector that defines the direction of the observer (see Section I of the Supplemental Material \cite{supp}).



The source terms for a Nambu-Goto string can be expressed using $X_{\pm}^{(1)}$ \cite{Burden:1985md, Battye:1993jv},
\begin{equation}
    \label{StrEnerTens}
    T^{\mu \nu}(x) = \frac{\mu}{2} \int  X^{(1)(\mu}_+ X^{(1)\nu)}_- \delta^{(4)}(x-X) \, \mathrm{d} \sigma_- \mathrm{d} \sigma_+,
\end{equation}
\begin{equation}
    \label{StrJTens}
    J^{\mu \nu}(x) = \frac{\eta}{2} \int  X^{(1)[\mu}_+ X^{(1)\nu]}_- \delta^{(4)}(x-X) \,  \mathrm{d} \sigma_- \mathrm{d} \sigma_+,
\end{equation}
where we highlight the different symmetrisations over the indices for the two cases. The current term $J^{\mu\nu}$ is relevant when the string is coupled to a massless scalar field 
for example when strings arise from the breaking of a global $U(1)$ symmetry. The Fourier transformed source terms can be split into components as follows,
\begin{equation}
\begin{gathered}
    \label{TJFourierTransform}
    \tilde{T}^{\mu \nu} = \mu I^{(\mu}_+ I^{\nu)}_-, \qquad    \tilde{J}^{\mu \nu} = \eta I^{[\mu}_+ I^{\nu]}_-,
\end{gathered}
\end{equation}
where
\begin{equation}
    \label{Ipm}
    I^{\mu}_{\pm} = \int^{\infty}_{-\infty} X_{\pm}^{(1)\mu} \text{e}^{-\frac{i}{2} \omega \left( \sigma_{\pm}-\hat{\textbf{n}}\cdot\textbf{X}_{\pm} \right) } \, \mathrm{d} \sigma_{\pm}.
\end{equation}
Thus, calculation of the spectrum of gravitational \cite{Vachaspati:1984gt,Burden:1985md,Garfinkle:1987yw, Sakellariadou:1990ne, Damour:2001bk, Binetruy:2009vt} and axion \cite{Vilenkin:1986ku,Garfinkle:1987yw,Sakellariadou:1991sd, Battye:1993jv} radiation from a Nambu-Goto string is reduced to the evaluation of \eqref{Ipm} for a chosen configuration.


In \cite{Damour:2001bk}, the massless emission from a cusp is calculated using the Taylor expansion \eqref{CuspTaylor} of the string solutions around the cusp point. Expression \eqref{Ipm} becomes
\begin{equation}
    \label{IpmAppr}
    I^{\mu}_{\pm} \approx \int^{\infty}_{-\infty} \sigma_{\pm} X_{\rm c \pm}^{(2)\mu} \text{e}^{-\frac{i}{2} \omega \sigma_{\pm} \left[1-\hat{\textbf{n}} \cdot \left( \textbf{X}^{(1)}_{\rm c \pm} +\textbf{X}^{(2)}_{\rm c \pm} \frac{\sigma_{\pm}}{2} + \textbf{X}^{(3)}_{\rm c \pm} \frac{\sigma^2_{\pm}}{6} \right) \right] }\, \mathrm{d} \sigma_{\pm},
\end{equation}
where terms proportional to $X_{\rm c \pm}^{(1)\mu}$ have been gauged away (see Section II of the Supplemental Material \cite{supp} for details) and the unit vector pointing toward the observer is denoted by $\hat{\textbf{n}}$. An analytic expression for the asymptotic behaviour when $\omega \rightarrow \infty$ can be obtained using the saddle point approximation \cite{Damour:2000gg} together with the cusp conditions \eqref{GenericCuspDiff}. Throughout this work, we consider the observer to be aligned with the direction of the maximum radiation burst, following the approach of \cite{Damour:2000wa}. In this case, $\hat{\textbf{n}} \cdot \textbf{X}^{(1)}_{{\rm c}\pm}=1$ and $\hat{\textbf{n}} \cdot \textbf{X}^{(2)}_{{\rm c}\pm}=0$, yielding
\begin{align}
\label{GenericCuspI}
    I^{\mu}_{\pm} &\approx \int^{\infty}_{-\infty} \sigma_{\pm} X^{(2)\mu}_{\rm c \pm} \text{e}^{ \pm \frac{i}{12} \omega \left( \textbf{X}^{(2)}_{\rm c \pm} \right)^2 \sigma_{\pm}^3 } \,\mathrm{d} \sigma_{\pm} \nonumber \\
    &= \pm i \frac{X^{(2)\mu}_{\rm c \pm}}{\left|\textbf{X}^{(2)}_{\rm c \pm}\right|^{4/3} } \left( \frac{12}{\omega} \right)^{2/3} \frac{\Gamma(\frac{2}{3})}{\sqrt{3}} \equiv \mathcal{I}^{\mu}_{\rm DV}.
\end{align}
Writing the waveform explicitly and neglecting the phase term, we obtain the asymptotic waveform
\begin{equation}
 \label{WaveformDV}
    \tilde{\kappa}^{\mu \nu}(\omega) \approx   \frac{\kappa_g \pi\mu}{\omega^\frac{4}{3}} \frac{4\pi (12)^{\frac{4}{3}}}{(3\Gamma(\frac{1}{3}))^2}\frac{X^{(2)\,(\mu}_{\rm c +}}{\left|\textbf{X}^{(2)}_{\rm c +}\right|^{4/3} }\frac{X^{(2)\,\nu)}_{\rm c -}}{\left|\textbf{X}^{(2)}_{\rm c -}\right|^{4/3} }.
\end{equation} 
Equation \eqref{WaveformDV} can be compared directly with equations (2.26) and (3.12) in \cite{Damour:2001bk}, yielding the asymptotic behaviour of the spectrum $\tilde{\kappa}^{\mu\nu} \propto \omega^{-4/3}$, where the direction of the radiation beam coincides with $\textbf{X}^{(1)}_{{\rm c}\pm}$. This calculation can also be applied to obtain the axion spectrum
\begin{equation}
 \label{AxionWaveformDV}
    \tilde{b}^{\mu \nu}(\omega) \approx   \frac{\kappa_a \pi \eta}{\omega^\frac{4}{3}} \frac{4\pi (12)^{\frac{4}{3}}}{(3\Gamma(\frac{1}{3}))^2}\frac{X^{(2)\,[\mu}_{\rm c +}}{\left|\textbf{X}^{(2)}_{\rm c +}\right|^{4/3} }\frac{X^{(2)\,\nu]}_{\rm c -}}{\left|\textbf{X}^{(2)}_{\rm c -}\right|^{4/3} }\,.  
\end{equation}




We can extend this method to study the radiation from the smooth cusp \eqref{Disc2Der}. Due to the discontinuous second derivative, we must split the integral \eqref{IpmAppr} into two parts either side of the discontinuity. We obtain
\begin{align}
    \label{SmoothI}
    I^{\mu}_{\pm} &\approx \int^{0}_{-\infty} |X^{(2)\mu}_{\text{c} \pm}| \sigma_{\pm} \text{e}^{\pm\frac{i}{12} \omega \left( \textbf{X}^{(2) }_{\text{c} \pm} \right)^2 \sigma^3_{\pm} }\, \mathrm{d}\sigma_{\pm} \nonumber \\ &\hspace{1cm} - \int^{\infty}_{0} |X^{(2) \mu}_{\text{c} \pm}| \sigma_{\pm} \text{e}^{\pm \frac{i}{12} \omega \left( \textbf{X}^{(2)}_{\text{c} \pm} \right)^2 \sigma^3_{\pm} }\, \mathrm{d}\sigma_{\pm} \nonumber \\
    &= - \frac{|X^{(2)\mu}_{\text{c} \pm}|}{\left| \textbf{X}^{(2)}_{\text{c} \pm} \right|^{4/3}} \left( \frac{12}{\omega  } \right)^{2/3} \frac{\Gamma(\frac{2}{3})}{3} \equiv \pm i\frac{\mathcal{I}^{\mu}_{\rm DV}}{\sqrt{3}},
\end{align}
where $|X^{(2)\mu}_{\text{c} \pm} | \equiv \lim_{\sigma \rightarrow  \sigma^+_{\rm c}} X^{(2)\mu}_{\text{c} \pm} (0,\sigma) $. The result \eqref{SmoothI} is smaller by a factor of $\sqrt{3}$ compared to the generic cusp expression \eqref{GenericCuspI}. Consequently, the radiated power has the same frequency dependence, but its amplitude is suppressed by a factor of $9$, assuming all other factors remain the same. A similar calculation can be performed for discontinuities in higher order derivative terms.



The expressions \eqref{GenericCuspI} and \eqref{SmoothI} are applicable whenever $\textbf{X}^{(2)}_{\text{c} \pm} \neq 0$. When $X^{(2) \mu}_{\text{c} \pm} = 0\,,$ we need to retain higher order terms of the Taylor expansion. The next non-trivial contribution is given by
\begin{align}
    \label{Integral4}
    I^{\mu}_{\pm} &\approx \frac{1}{2} \int^{\infty}_{-\infty} X^{(3)\mu}_{\text{c} \pm} \sigma^2_{\pm} \text{e}^{-\frac{i}{2} \frac{3}{5!} \omega \left( \textbf{X}^{(3)}_{\text{c} \pm} \right)^2 \sigma^5_{\pm} } \, \mathrm{d} \sigma_{\pm} \nonumber \\
    &= \frac{X^{(3)\mu}_{\rm c \pm}}{5} \left( \frac{2 \cdot 5!}{3 \omega\, (\textbf{X}^{(3)}_{\rm c \pm})^2} \right)^{3/5} \cos\left( \frac{3\pi}{10}\right) \Gamma\left(\frac{3}{5}\right).
\end{align}
In this case, we obtain $I^{\mu}_{\pm} \propto \omega^{-3/5}$ in contrast to $I^{\mu}_{\pm} \propto \omega^{-2/3}$ in \eqref{GenericCuspI} for a generic cusp. For general $\textbf{X}^{(n)}_{\rm c \pm} \neq 0$ with $\textbf{X}^{(p)}_{{\rm c} \pm} = 0$ for all $1 < p < n$, including cases with discontinuous derivatives at the cusp point, an explicit example of which is given by \eqref{CrazyCusp}, we obtain
\begin{equation}
\begin{gathered}
    \label{Igeneric}
    I^{\mu}_{\pm} \approx \mathcal{I}^{\mu}+(-1)^{n+1} \left(\mathcal{I}^{\mu} \right)^* ,
\end{gathered}
\end{equation}
where
\begin{align}
\label{UsefulIntegral}
\mathcal{I}^{\mu} &=\int^{\infty}_{0}  \frac{\sigma^{n-1}_{\pm} X^{(n) \mu}_{\rm c \pm}}{(n-1)!} \exp\left[\pm   \frac{i \omega \left(\textbf{X}^{(n)}_{\rm c \pm}\right)^2 \sigma^{2n-1}_{\pm}}{4 (2n-1) \left[(n-1)! \right]^2}  \right] \mathrm{d} \sigma_{\pm} \nonumber \\
    &= \left( \frac{ 4 n \left[(n-1)! \right]^2 }{ \beta_n \omega \left( \textbf{X}^{(n)}_{\rm c \pm} \right)^2} \right)^{\beta_n} \frac{ \Gamma \left( \beta_n \right)  X^{(n) \mu}_{\rm c \pm} }{(2n-1) (n-1)!} \text{e}^{\pm i \beta_n \frac{\pi}{2} }
\end{align}
and $\beta_n = \frac{n}{2n-1}$ where $n\geq2$. 
This leads to the general frequency dependence $|I^{\mu}_{\pm}| \propto \omega^{-\beta_n}$. Since the left- and right-moving modes are independent, the waveform of gravitational or axion emission for $\omega \rightarrow \infty$ takes the form
\begin{equation}
    \label{NewSpectr}
    \tilde{\kappa}^{\mu \nu}(\omega)\, , \tilde{b}^{\mu \nu}(\omega) \propto \omega^{-(\beta_n+\beta_m) }, \quad n,m\geq2 ,\quad n,m\in\mathbb{Z},
\end{equation}
where $n$ and $m$ denote the lowest non-zero derivative for each mode. 

Taking the limit $\lim_{n \rightarrow \infty} \beta_n = \frac{1}{2}$ implies that the slowest possible spectral decay is $\tilde{\kappa}^{\mu \nu} \propto \omega^{-1}$. This corresponds to the unphysical situation of a straight string moving at the speed of light, as can be seen from taking the limit of \eqref{CrazyCusp} as $n\rightarrow \infty$. Simiarly, for circular loops, all points on the string move at the speed of light at the moment of collapse to a point, leading again to $\tilde{\kappa}^{\mu \nu} \propto \omega^{-1}$. This has been calculated analytically for Nambu-Goto strings \cite{Battye:1993jv, Dine:2020pds, Rybak:2022sbo} and is in agreement with the measured spectrum from loops in field theory simulations \cite{Hagmann:1990mj,Benabou:2023ghl}. In this case, we do not expect a beamed signal, because all points of the string are a cusp point. Finally, we note that the spectrum for $n=1$ corresponds to the predicted frequency dependence $\tilde{\kappa}^{\mu \nu} \propto \omega^{-2}$ of a signal from a kink-kink collision \cite{Damour:2001bk, Binetruy:2009vt}.

\textit{More Realistic Cusps}---For cosmic strings arising from spontaneously broken symmetries, the Nambu-Goto action is the leading-order term in an effective field theory (EFT) describing the underlying `real' field theory string. Higher-order corrections, for example, from the finite string thickness or from backreaction effects \cite{Battye:1995hw, Anderson:1997ip, Buonanno:1998is, Auclair:2021jud, Blanco-Pillado:2023sap,  Wachter:2024aos, Wachter:2024zly, Aurrekoetxea:2026xrl} can modify the string dynamics and may prevent the formation of cusps. 
For cosmic superstrings, although the zero-width approximation is exact \cite{Copeland:2003bj}, additional effects, such as the presence of extra dimensions, can obstruct cusp formation \cite{OCallaghan:2010mtk}. In these cases, an exact overlap of separable left- and right-moving modes on the Kibble-Turok sphere will generally fail to occur, and the resulting massless spectrum emitted from a cusp-like region will exhibit slight deviations from that predicted in the strict Nambu-Goto limit.


We can allow for small departures from the exact Nambu-Goto solutions by introducing approximate solutions $\bar{\mathbf{X}}^{(1)}_\pm$, moving `off' the Kibble-Turok sphere by some small amount. These take the form
\begin{equation}
\label{CuspDevi}
\left|\bar{\textbf{X}}^{(1)}_{\pm}\right|^2 = 1 - f_{\pm}(\delta_\pm) \approx 1 - \delta_\pm f_{\pm}'(0) \equiv 1 - 2 \Delta_{\pm}, \quad \Delta_
\pm<<1\,,
\end{equation}
where $f_\pm$ are arbitrary functions of some small corrections $\delta_\pm$. The approximate left- and right-moving solutions no longer intersect precisely, but instead approach each other to within a small interval. The $\pm$ modes now propagate at velocities slightly below the speed of light.

In the previous analysis, the cusp conditions \eqref{GenericCuspDiff} ensure that the terms $\hat{\mathbf{n}}\cdot\mathbf{X}_\pm^{(2)}$ in the Taylor expansion vanish at the cusp point. When the cusp conditions are no longer satisfied exactly, lower-order terms are retained, giving
\begin{equation}
    \label{ThickCusp}
    I^{\mu}_{\pm} \approx \int_{-\infty}^{\infty} \sigma_{\pm} \bar{X}^{(2) \mu}_{\rm c \pm} \text{e}^{- \frac{i}{2} \omega \sigma_{\pm} (\Delta_{\pm} + \beta_{\pm} \sigma_{\pm} + \gamma_{\pm} \sigma_{\pm}^2)  }\, \mathrm{d} \sigma_{\pm}\,,
\end{equation}
where $\Delta_{\pm} \approx 1 - \hat{\textbf{n}} \cdot \bar{\textbf{X}}^{(1)}_{\rm c \pm}$, $\beta_{\pm} \equiv \hat{\textbf{n}} \cdot \bar{\textbf{X}}^{(2)}_{\rm c \pm} $ and $\gamma_{\pm} \equiv \hat{\textbf{n}} \cdot \bar{\textbf{X}}^{(3)}_{\rm c \pm}$. This integral can be solved exactly \cite{Rybak:2022sbo} to obtain 
\begin{equation}\label{highfcutoff}
   I^{\mu}_{\pm} = 2 \pi i \Phi 
\frac{ \bar{X}^{(2) \mu}_{\rm c \pm} \sgn(\gamma_{\pm})  }{3^{2/3} \omega^{2/3} \left| \gamma_{\pm} \right|^{2/3}}  
\left( 
\frac{i \beta_{\pm} \, \omega^{1/3}}{ ( 3 \left| \gamma_{\pm} \right| )^{2/3}} \, \mathrm{Ai}(\zeta_{\pm}) 
+ \mathrm{Ai}'(\zeta_{\pm}) 
\right),
\end{equation}
where $\zeta_{\pm} = - \frac{(\beta_{\pm}^2 - 3 \Delta_{\pm} \gamma_{\pm}) \omega^{2/3}}{3^{4/3} \left| \gamma_{\pm} \right|^{4/3}}$, $\Phi_{\pm} = \text{e}^{ - i \beta_{\pm} \omega \frac{ 2 \beta_{\pm}^2 - 9 \Delta_{\pm} \gamma_{\pm} }{27 \gamma_{\pm}^2} }$ and $\mathrm{Ai}(\dots)$ and $\mathrm{Ai}'(\dots)$ are the Airy function and its derivative, respectively. Setting $\Delta_{\pm}= \beta_{\pm} =0$, we recover the result for generic cusps, since $\mathrm{Ai}'(0) = - \frac{ 3^{-1/3} }{ \Gamma(1/3)}$. 

\begin{figure}[h!]
	\centering 
\includegraphics[width=0.48\textwidth, angle=0]{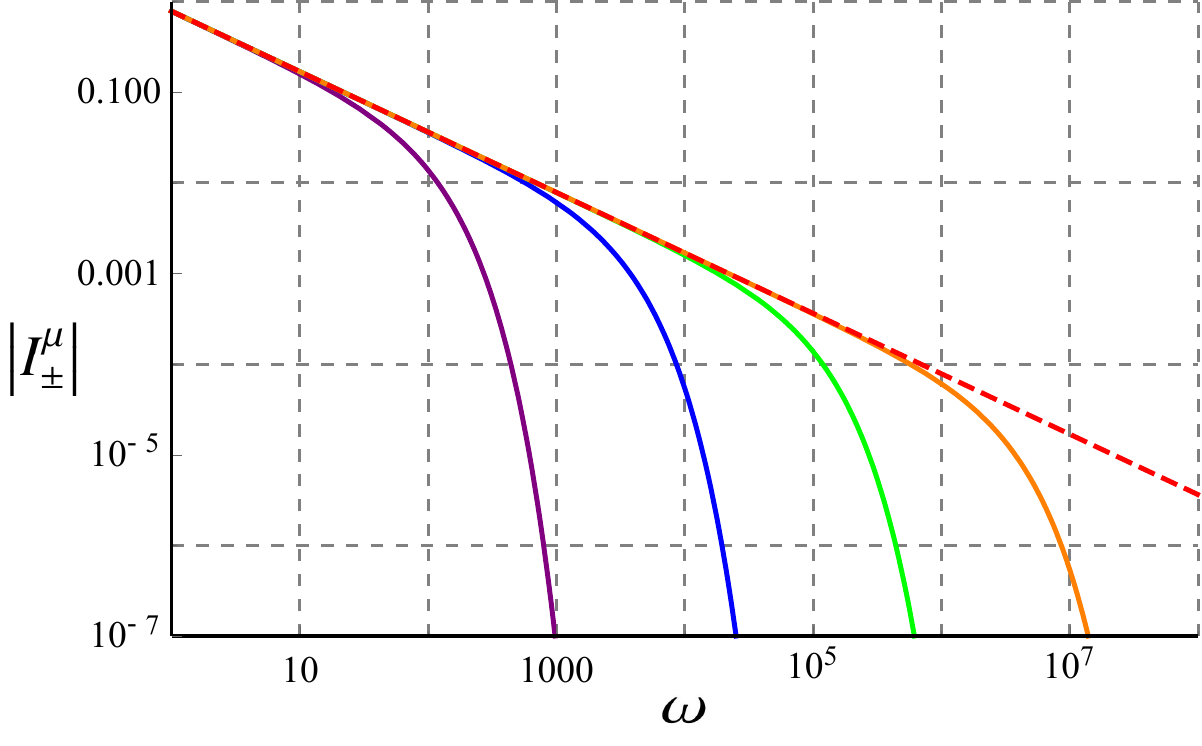}\\
\caption{Radiation spectra computed using \eqref{highfcutoff}. Lines for $\Delta_{\pm} = 10^{-1}$ to $10^{-4}$ are compared to the dashed $|I^\mu_\pm|\propto \omega^{-2/3}$ power law. Larger $\Delta_{\pm}$ shift the cutoff to lower frequencies.} 
	\label{Fig:Cusp5}%
\end{figure}

Figure~\ref{Fig:Cusp5} illustrates how deviations $\Delta_
\pm$ from the Nambu–Goto approximation modify the components $|I^\mu_\pm|$ of the massless spectrum, introducing a high-frequency cutoff to the $|I^\mu_\pm| \propto \omega^{-2/3}$ power-law behaviour. Smaller corrections come from changing $\beta_\pm$ and $\gamma_\pm$. 

\textit{Discussion}---In the canonical Nambu-Goto picture, cosmic string cusps are characterised as points on a string loop that arise instantaneously, move at the speed of light, generate massless radiation with a $\tilde{\kappa}(\omega) \sim \omega^{-4/3}$ power law spectrum and have a sharp shape $y \sim x^{2/3}$. We have broadened this picture by identifying and computing the shapes and massless spectra of several previously unexplored cusp configurations. We have found that some formal cusps exhibit sharper profiles than the standard $y \sim x^{2/3}$, while others are smooth. The cusp shape does not directly determine the massless spectrum; in particular, smooth cusps can reproduce the standard $\tilde{\kappa}(\omega) \sim \omega^{-4/3}$. The novel spectra derived fill the range between $\tilde{\kappa}(\omega) \sim\omega^{-4/3}$ and $\omega^{-1}$ with discrete power laws \eqref{NewSpectr}, revealing a far richer structure of possible signatures than previously recognised. The non-generic cusps introduced may not frequently arise in cosmological networks, as the requirement for higher derivatives to identically vanish imposes fine-tuning constraints on the string worldsheet. However, they could potentially arise with statistical significance in cosmological scenarios featuring exceptionally large loop populations, such as superheated phase transitions (e.g. \cite{East:2022rsi}) or certain models of high-tension networks. Fully quantifying their abundance in these scenarios is left for future work.





We highlight that, in addition to the usual sharp cusps, smooth cusp configurations involving true discontinuities also receive (infinitely) large corrections from higher derivative terms in the string curvature. Here, one expects excitations of the radial mode to propagate along the string, as demonstrated numerically in \cite{Drew:2023ptp}. This highlights that deviations from Nambu-Goto behaviour in field theory strings need not be confined to sharply localised features. Backreaction will similarly smooth the pointed cusp shape and reduce the overall strength of the signal \cite{Wachter:2024aos, Wachter:2024zly}. For the already smooth cusps investigated here, we expect that exact cusp formation will be prevented by the reduction of the local velocity. 
In all cases, such corrections will lead to a high frequency cutoff in the signal associated with the energy or length scale of the relevant effect, modifying the predicted spectrum both for individual bursts and stochastic background signals from a string network \cite{LIGOScientific:2017ikf, LIGOScientific:2021nrg, Auclair:2019wcv}. This should be taken into account in particular for future experimental GW burst searches, for which the cutoff could be used to infer information about the physics of the string core. Further investigation is left for future work.

Finally, this work opens a potential avenue for future simulations of cosmic string cusps based on the collision of travelling waves, following from those performed in \cite{Drew:2023ptp}. These could be used to verify the massless spectra for the new cusps presented, including the dependence on string parameters of the position of the high frequency cutoff that comes from corrections to pure Nambu-Goto.


\textit{Acknowledgements}---We thank E.P.S. Shellard, J.J. Blanco-Pillado, S. Dutta Chowdhury, D. Cyncynates, N.A. Dondi, T. Jain, M. Kaltschmidt, M. Mirbabayi, J. Redondo, G. Villadoro and J. Wachter for helpful discussions. AD was supported by funding from the European Union’s Horizon Europe research and innovation programme under the Marie Skłodowska-Curie grant agreement No 101151409 (GWStrings), and acknowledges support from a Junior Research Fellowship (JRF) at Homerton College, University of Cambridge. Views and opinions expressed are however those of the authors only and do not necessarily reflect those of the European Union or the European Research Executive Agency (REA). Neither the European Union nor the granting authority can be held responsible for them. IR is supported by IBS under the project code IBS-R018-D3. This article is based upon work from COST Action COSMIC WISPers CA21106, supported by COST (European Cooperation in Science and Technology). This work was done in the context of FCT project 2022.04048.PTDC (Phi in the Sky, DOI 10.54499/2022.04048.PTDC). For open access purposes, AD has applied a CC BY public copyright license to any author-accepted manuscript version arising from this submission. 

\begin{figure}[h!]
    \centering
    \includegraphics[width=\linewidth, trim = 0 0 0 .5cm, clip]{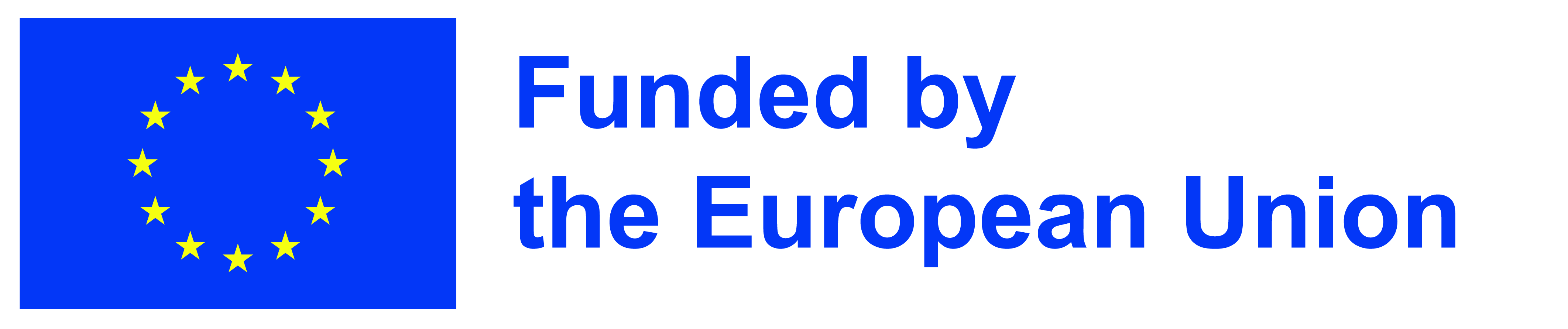}
\end{figure}

\vspace{-.5cm}

\bibliographystyle{apsrev4-2}
\bibliography{sample}

\end{document}


\title{Supplemental Material for `LQ20139DR: New massless spectra from cosmic string cusps'}

\maketitle

\preprint{APS/123-QED}

\author{Amelia Drew}
\affiliation{The Abdus Salam ICTP, Strada Costiera 11, 34151, Trieste, Italy}
\affiliation{Institute for Fundamental Physics of the Universe, Via Beirut 2, 34151 Trieste, Italy}

\author{Ivan Rybak}
\affiliation{Cosmology, Gravity, and Astroparticle Physics Group, Center for Theoretical Physics of the Universe, Institute for Basic Science (IBS), Daejeon, 34126, Korea}
\affiliation{CAPA \& Departamento de F\'{\i}sica Te\'{o}rica, Universidad de Zaragoza, Pedro Cerbuna, 12, Zaragoza, 50009}

\date{\today}

\appendix

\section{Section I: Waveform in the Local Wave Zone for Non-periodic Source}\label{AppA}

In this Appendix, we provide details of the derivation of expression (22). While the derivation is very similar to the one presented in Section II.A of \cite{Damour:2001bk} for a periodic source, we include the details for a non-periodic source for the sake of completeness.

The field equations for the gravitational and Kalb-Ramond cases take a similar form (21), allowing us to obtain the radiation waveform expression for both cases simultaneously. Thus, we consider
\begin{equation}
\label{EqOfMotApp}
    \Box \varphi^{\mu \nu}(t,\textbf{x}) = 4 \pi \kappa S^{\mu \nu} (t,\textbf{x}),
\end{equation}
where $\varphi^{\mu \nu}$ represents the gravitational field $\bar{h}^{\mu \nu}$ or the Kalb-Ramond field $B^{\mu \nu}$, and $S^{\mu \nu}$ is a source with the corresponding constant $\kappa$, encapsulating the equations (21). We represent the source and the field by the Fourier transform:
\begin{equation}
\begin{gathered}
\label{FourierTrSPhi}
    S^{\mu \nu}(t,\textbf{x}) \equiv \frac{1}{2 \pi} \int^{\infty}_{- \infty} \text{e}^{-i \omega t} \tilde{S}^{\mu \nu}(\omega,\textbf{x}) d \omega, \\
    \varphi^{\mu \nu}(t,\textbf{x}) \equiv \frac{1}{2 \pi} \int^{\infty}_{- \infty} \text{e}^{-i \omega t} \tilde{\varphi}^{\mu \nu }(\omega,\textbf{x}) d \omega.
\end{gathered}
\end{equation}
Substituting \eqref{FourierTrSPhi} into \eqref{EqOfMotApp} leads to the Helmholtz equation:
\begin{equation}
    \label{Helm}
    \left( \Delta + \omega^2 \right) \tilde{\varphi}^{\mu \nu}(\omega,\textbf{x}) = - 4 \pi \kappa \tilde{S}^{\mu \nu} (\omega,\textbf{x}).
\end{equation}
The solution of equation \eqref{Helm}, found using the Green's function, takes the form
\begin{equation}
    \label{PhiSol}
    \tilde{\varphi}^{\mu \nu}(\omega,\textbf{x}) = \kappa \int \frac{\text{e}^{i \omega | \textbf{x}-\textbf{x}' | }}{| \textbf{x}-\textbf{x}' |} \tilde{S}^{\mu \nu}(\omega,\textbf{x}') d^3 x
\end{equation}
using the approximation that the source—the location of the travelling wave collision—is localised at $\textbf{x}'=0$ and is much smaller than the distance $r$ to the observer. In that case, we can approximate the solution as
\begin{equation}
\begin{gathered}
    \label{PhiSol2}
    \tilde{\varphi}^{\mu \nu}(\omega,\textbf{x}) \approx \kappa \int \frac{\text{e}^{i \omega (r - \hat{\textbf{n}} \cdot \textbf{x}' ) }}{r} \tilde{S}^{\mu \nu}(\omega,\textbf{x}') d^3 x = \\
    = \kappa \frac{\text{e}^{i \omega r }}{r} \tilde{S}^{\mu \nu}(\omega,\hat{\textbf{n}}),
    \end{gathered}
\end{equation}
where $S^{\mu \nu} (\omega,\hat{\textbf{n}}) \equiv \int \text{e}^{i k^{\rho} x_{\rho}} S^{\mu \nu}(t,\textbf{x}) d^4 x$ is the spacetime Fourier transform of the source, and $\hat{\textbf{n}}$ is a unit vector that defines the direction of the observer. Hence, defining the distance-independent asymptotic waveform, the time-Fourier-transformed solution for the field can be written as
\begin{equation}
    \tilde{\phi}^{\mu \nu} (\omega,\hat{\textbf{n}}) \equiv \tilde{\varphi}^{\mu \nu} (\omega,\textbf{x}) r \approx \kappa \text{e}^{i \omega r} \tilde{S}^{\mu \nu}(\omega,\hat{\textbf{n}}),
\end{equation}
which are the expressions given in (22). \\

\section{Section II: Dominant Terms for Gravitational and Axion Radiation} \label{AppB}

In this Appendix, we explicitly demonstrate how gauge degrees of freedom can be used to eliminate unphysical contributions to radiation arising from terms  $\propto X^{(1)}_{\pm \mu} $ \cite{Damour:2001bk}. 

The linearised coordinate transformations in Fourier space for the gravitational field lead to
\begin{equation}
\label{grGauge}
    \tilde{\kappa}'_{\mu \nu} \rightarrow \tilde{\kappa}_{\mu \nu} + k_{\mu} \xi_{\nu} + k_{\nu} \xi_{\mu},
\end{equation}
while gauge transformations in Fourier space for the Kalb-Ramond field require
\begin{equation}
\label{axGauge}
    \tilde{b}'_{\mu \nu} \rightarrow \tilde{b}_{\mu \nu} + k_{\mu} \xi_{\nu}-k_{\nu} \xi_{\mu},
\end{equation}
where $k_{\mu} \propto X^{(1)}_{\mu}$ is a light-like vector.

Using (22) together with (26), one finds that the asymptotic, distance-independent waveforms for the gravitational and axion fields are given by
\begin{equation}
    \begin{gathered}
    \label{GrAxProp}
        \tilde{\kappa}_{\mu \nu} \propto \left(a_+k_{(\mu} + b_+  X^{(2)}_{+(\mu} \right) \left( a_- k_{\nu)} + b_- X^{(2)}_{-\nu)}  \right),\\
        \tilde{b}_{\mu \nu} \propto \left(a_+k_{[\mu} + b_+ X^{(2)}_{+[\mu} \right) \left( a_- k_{\nu]} + b_- X^{(2)}_{-\nu]}  \right),
    \end{gathered}
\end{equation}
By choosing the vector 
\begin{equation}
  \xi_{\mu} \propto  k_\mu + \frac{b_+}{a_+} X^{(2)}_{+ \mu} + \frac{b_-}{a_-} X^{(2)}_{- \mu} , 
\end{equation}
in the gauge transformations  \eqref{grGauge} and \eqref{axGauge}, we can explicitly cancel all terms in \eqref{GrAxProp} that are $ \propto k_\mu k_\nu$, $\propto k_{\mu}X^{(2)}_{+\nu}$, and $k_{\mu}X^{(2)}_{-\nu}$. This procedure isolates the physical contribution, which is solely $\propto X^{(2)}_{+\mu} X^{(2)}_{-\nu}$, thereby justifying the form of expression (28).

\bibliographystyle{apsrev4-2}
\bibliography{supplemental}